\documentclass[a4paper,11pt]{article}
\usepackage{pos}

\title{Untangling the Complexity in the Galactic Centre: a way to understand the origin of the gamma-ray emission from the inner Galaxy}
 \ShortTitle{Untangling the Complexity in the Galactic Centre}
\author*[a]{Sofia Ventura}

\affiliation[a]{Universit\`a  di Siena and INFN Pisa,\\
  I-53100 Siena, Italy}

\emailAdd{sofia.ventura@pi.infn.it}

\abstract{The origin of the high-energy gamma-ray emission from the Milky Way center is still unclear and debated because of the impact of systematics afflicting the measurements from current experiments. Several theories and phenomenological models attempt to explain the intricate panorama. The presence of a {\it PeVatron} in the Central Molecular Zone or in its vicinity, the contribution of the {\it hard-component} of the diffuse gamma-ray emission, and dark matter annihilation scenario are among the most promising mechanisms for describing the observed excess. The development of increasingly precise models able to reproduce the measured gamma-ray emission is the challenge for the scientific community in view of the next generation telescopes.

A detailed treatment of phenomenological models for the dubbed Cosmic Rays {\it Sea} (CR-{\it sea}) characterized by different configurations is scrutinized in comparison with the observed spectrum in the inner Galaxy, using \texttt{DRAGON} and \texttt{GAMMASKY} codes.}

\FullConference{37$^{\rm{th}}$ International Cosmic Ray Conference (ICRC 2021)\\
		July 12th -- 23rd, 2021\\
		Online -- Berlin, Germany}

\begin{document}
\maketitle

\section{Introduction}
Our Own Galaxy is known to be a strong source of diffuse $\gamma$-ray emission \citep{1975ApJ...198..163F,1972ApJ...177..341K}. First EGRET instrument  measured its flux over the full sky ( $E=0.03\div10$ GeV, \cite{1997ApJ...481..205H}). The next generation telescope {\it Fermi}-LAT -- onboard the NASA GLAST satellite -- measured with better accuracy the diffuse emission from the whole Galactic Plane (GP) up to than $100\ \rm{GeV}$ \citep{2012ApJ...750....3A}.

The most significant component of the diffuse emission is due to the decay of neutral pions ($\pi^0$), as consequence of the collisions of the Galactic Cosmic Rays (CRs) hadronic component with the gaseous structure of the Interstellar Medium (ISM, \cite{1970Ap&SS...6..377S,1986A&A...157..223D}): the column density of the interstellar matter is traced by the pionic flavour of the Galactic diffuse $\gamma$-ray emission. 
The leptonic component of CRs contributes to this emission  via Inverse Compton (IC) scattering and bremsstrahlung, but with different spectral shape and spatial distribution allowing to disentangle its effect to the measured flux. 

The Imaging Cherenkov Atmospheric Technique (IACT) telescopes observe the Very High Energy ($E>100\ \rm{GeV}$, VHE) $\gamma$-rays from the Earth ground \citep{1997APh.....6..343A,1997APh.....6..369A}: the restricted field-of-view (FOV) of these telescopes allows to study the diffuse emission in specific regions of interest, but with much higher angular resolution, and increasing the reachable energy range with respect to orbital observatories -- like {\it Fermi}-LAT characterized by a lower energy sensitivity, but a broader FOV.

Understanding and modeling the $\gamma$-ray diffuse emission, especially at higher energies, is fundamental  in the High Energy (HE) Astrophysics community: the capability to describe and reproduce the CR transport within Our Galaxy, and the subsequently interactions with the interstellar gas -- locked in complex structures -- gives the opportunity to estimate the expected emission in each part of the Milky Way. The reproducibility of the so-called large-scale CR-{\it sea} is an important aspect in the current, and especially, next generation telescopes since it has a key-role in the definition of the {\it background} model, a fundamental ingredient in the analysis-chain of measured data.

\section{The Central Molecular Zone}
The Galactic Center (GC) of the Milky Way  is the perfect laboratory for studying phenomena and physical processes that may be occurring in many other galactic nuclei. 
The observations show the evidence for a central massive black hole -- the compact radio source Sgr A$^\star$ -- and  a dense and luminous star cluster, as well as several components of neutral, ionized, and hot gas \citep{2010RvMP...82.3121G}. 
The inner part of the GC is represented by the so-called {\it Central Molecular Zone} (CMZ) which represents one of the densest environment of the Galaxy.

The first measurements of energy spectrum and spatial distribution of the diffuse $\gamma$-ray emission from the GC region was provided by the H.E.S.S. collaboration in $2006$ \citep{2006Natur.439..695A}. 
That emission was found tracing the column density of the interstellar gas (probed by the related CO and CS molecular emission  lines) and extends over about $2$ degrees in Galactic longitude along the GP: namely the {\it Central Galactic Ridge} region. 
This is the first evidence of an harder spectrum than that measured at the Earth position\footnote{It was expected to be the same in the whole Galaxy.}.

In $2016$ the H.E.S.S. collaboration released new data with much higher statistics ($\sim250$ hours of exposure) allowing to extend the explored energy range, and to perform a better study of the morphology of the emission \citep{2016Natur.531..476H, 2018A&A...612A...9H}. 
This flux was supposed to be originated by the interaction of high energy CR hadrons (mostly protons) with the dense gas in the GC region giving rise to $\pi^0$s, which rapidly decay into $\gamma$-rays. 
Assuming that the emission is originated by {\it proton-proton} scattering, this implies the presence of a population of primary CRs in that region extending up to energies close to $\sim1\ \rm{PeV}$ with a power-law (PL) spectrum with index close to $\sim2.4$ due to the slowly increasing behaviour of the $pp$-scattering cross section with energy (see \cite{Patrignani:2016xqp}). 
The presence of strong magnetic fields ($\sim100\ \mu\rm{G}$ \cite{2010Natur.463...65C}), and intense infrared (IR) radiation field produces relevant synchrotron and IC losses above the TeVs: the CR leptonic component is not expected to contribute significantly to the observed emission. 

The $\gamma$-ray diffuse emission traces the whole molecular gas complex in the GC (the CMZ), which approximates a cylindrical region of $\sim250\ \rm{pc}$ radius from the center. 
The inner part of the CMZ (the {\it Galactic Ridge}, $|l|<1^\circ$, $|b|<0.3^\circ$) {\footnote{$1^\circ$ corresponds to $\sim 150$ pc at the GC distance ($\sim 8.5\ \rm{kpc}$).}}, approximately corresponds to a symmetric cylinder of $\sim30\ \rm{pc}$ thickness and $\simeq 170$ pc long, around Sgr A$^\star$. 
The H.E.S.S. collaboration \citep{2018A&A...612A...9H} measured a uniform spectral index  $\Gamma_{\rm HESS18} = - 2.28 \pm 0.03_\text{stat} \pm 0.2_\text{syst}$ up to $45\ \rm{TeV}$, suggesting that a single population of particles fills the whole CMZ. 

The H.E.S.S. results have raised a wide interest in the Astroparticle community as they may provide the first evidence of a \emph{PeVatron} in the Milky Way.   

\section{Background Models for the CMZ}
\begin{figure}[ht] 
  \begin{minipage}[b]{0.5\linewidth}
    \includegraphics[width=1.\linewidth]{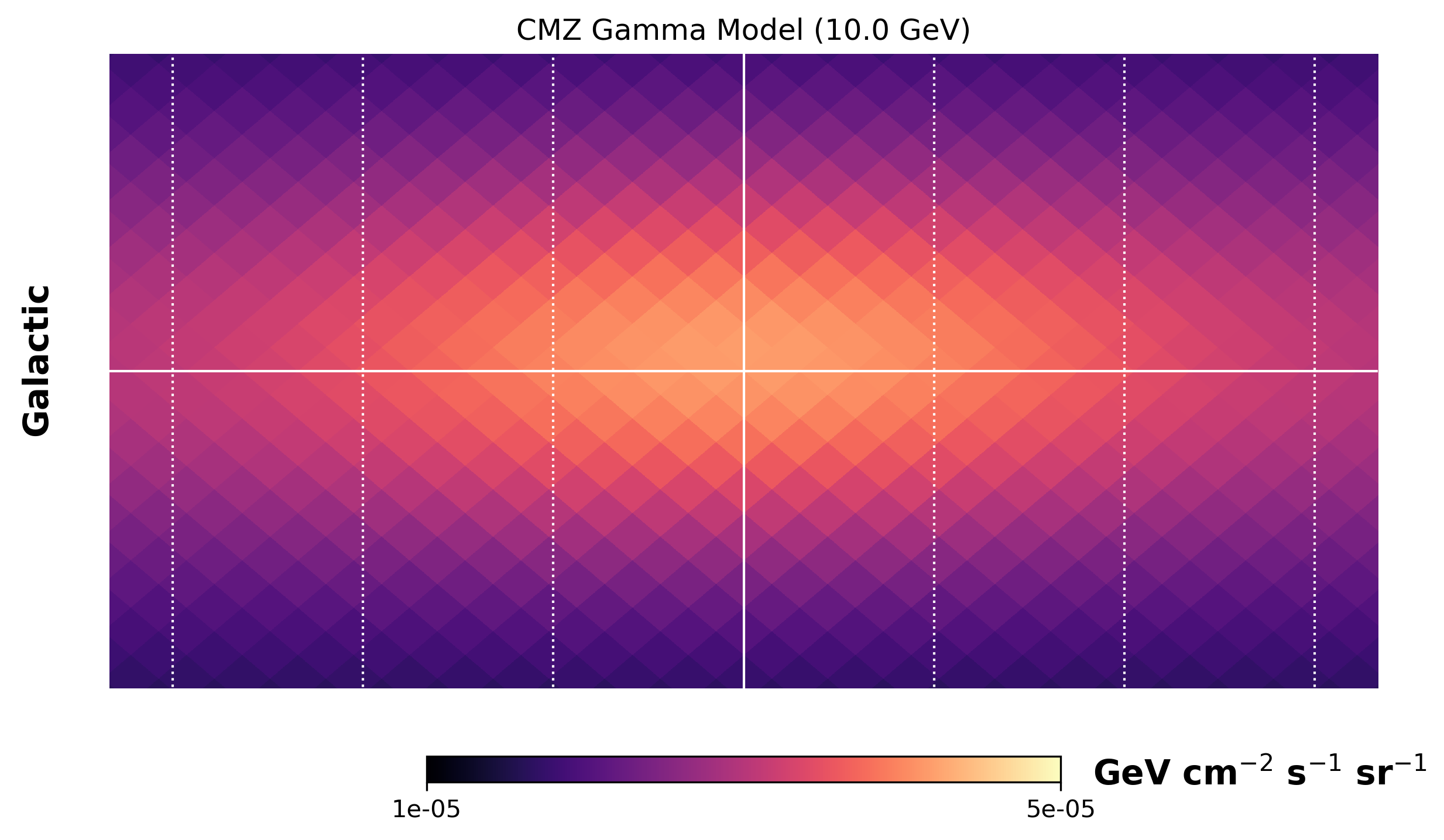} 
  \end{minipage} 
  \begin{minipage}[b]{0.5\linewidth}
    \includegraphics[width=1.\linewidth]{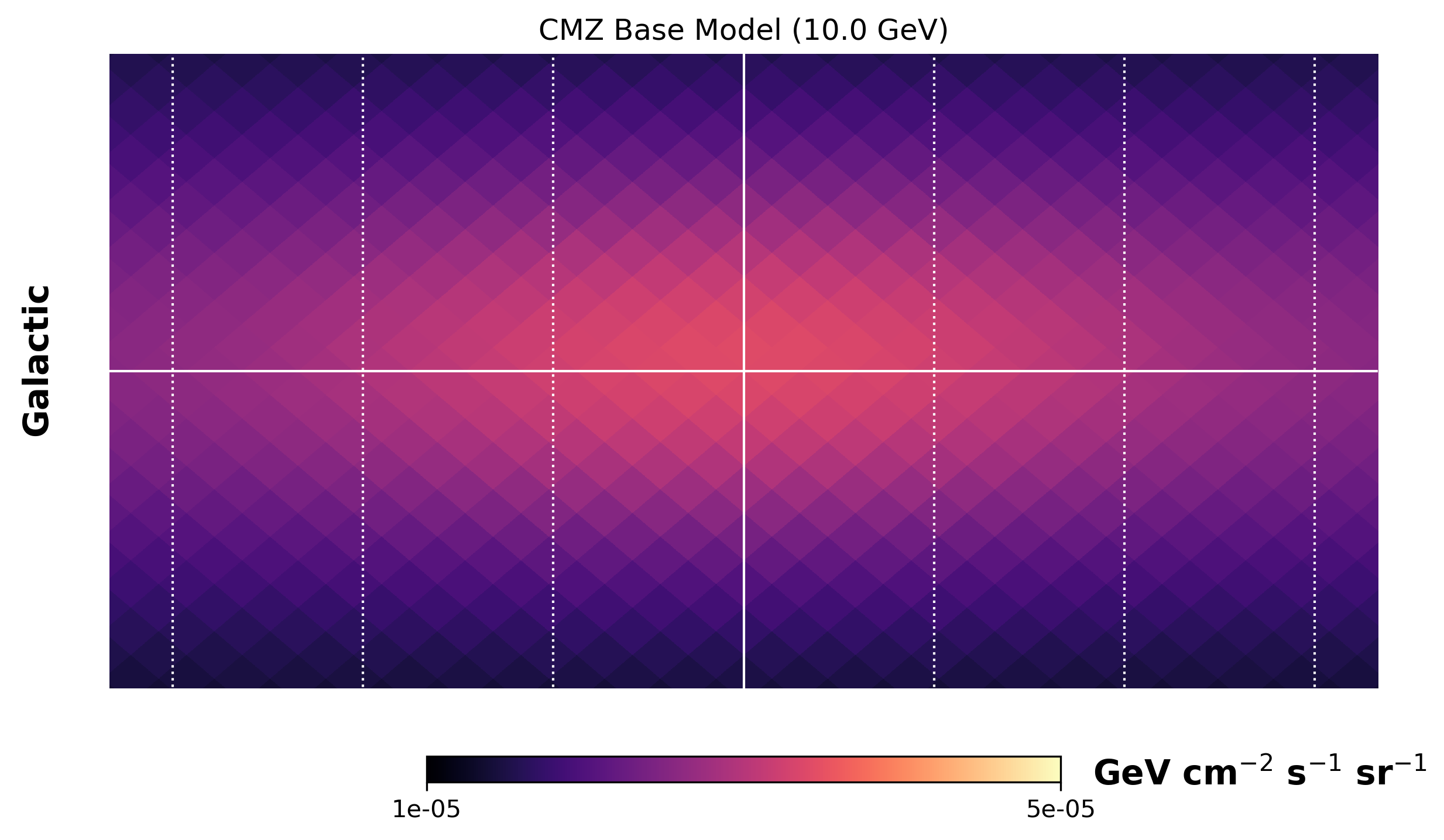} 
  \end{minipage} 
  \begin{minipage}[b]{0.5\linewidth}
    
    \includegraphics[width=1.\linewidth]{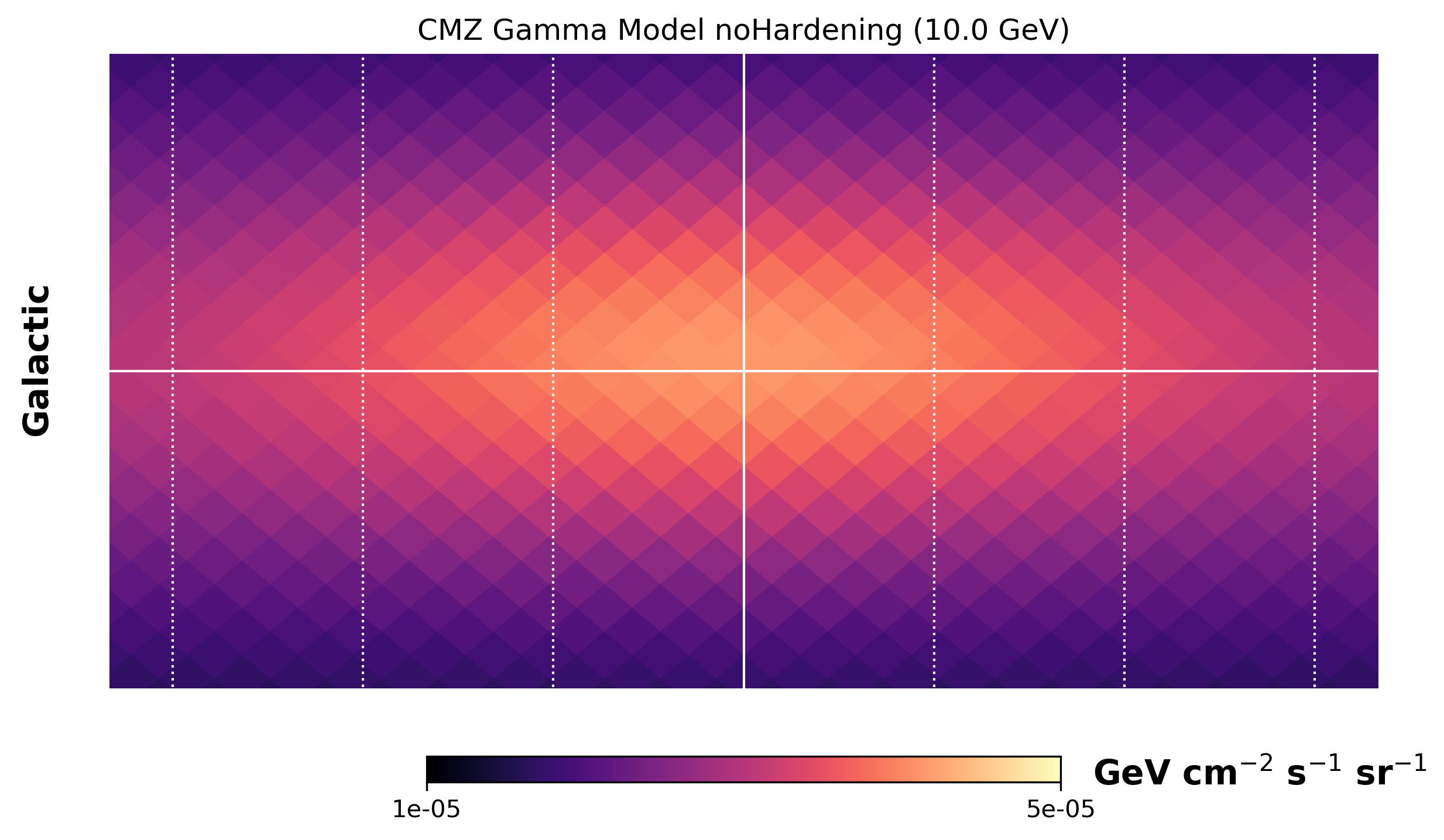}

  \end{minipage}
  \hfill
  \begin{minipage}[b]{0.5\linewidth}
    \includegraphics[width=1.\linewidth]{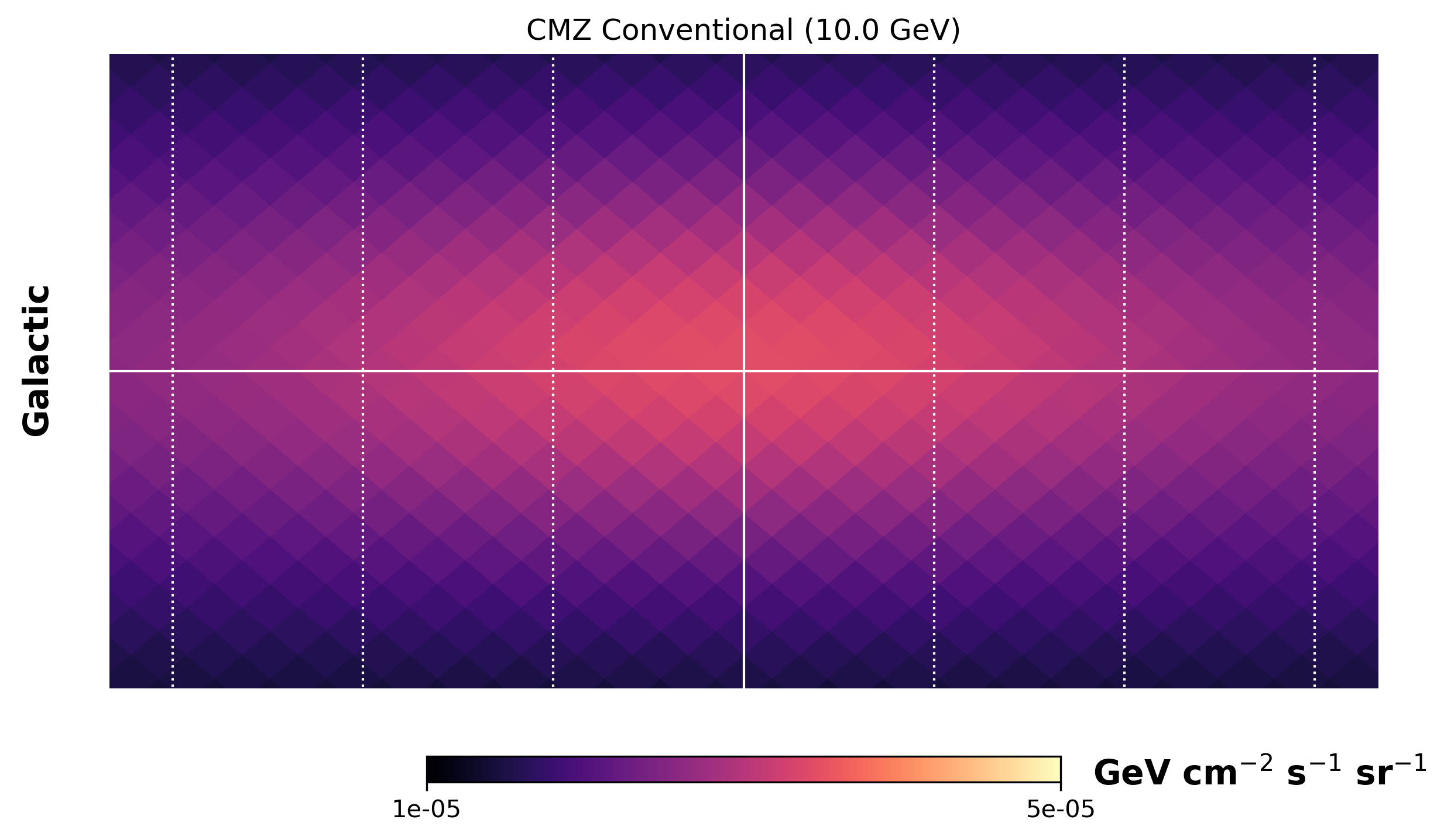} 
  \end{minipage} 
  \caption{View of the $\gamma$-ray expected emission as output of {\it Gamma Model}, {\it Gamma Model without hardening}, {\it Base Model} and {\it Conventional} description. The representations are obtained at $10\ \rm{GeV}$, and in a region corresponding to the Galactic Ridge ($|l|<1^\circ,\ |b|<0.3^\circ$).  }
\label{modview} 
\end{figure}

A nowadays important research field is represented by the efforts of the scientific community in modelling the large-scale $\gamma$-ray background. 
In figure \ref{modview} are reported four views of the  expected emission from the CMZ at $10\ \rm{GeV}$. Starting from the top left, {\it Gamma Model}, {\it Gamma Model without hardening}, {\it Base Model}, and {\it Conventional} description are shown. In this work is briefly described the characteristics of  such models.
 %
%

The so-called {\it Gamma Model} is obtained as the solution of transport equation -- solved by \texttt{DRAGON} code \citep{2008JCAP...10..018E} -- and integration along the line-of-sight of the CR density propagated in the whole Galaxy (using \texttt{GAMMASKY} code \citep{2013JCAP...03..036D,2012PhRvL.108u1102E}).  
This model is featured by the radial dependence of both the scaling of the CR diffusion coefficient with rigidity, and the advection velocity (for major details see \cite{2015JCAP...12..056G}).  This feature is able to reproduce the {\it hardening} of the CR spectral index observed in the inner Galaxy (see \cite{2016ApJS..223...26A, 2016PhRvD..93l3007Y}). 
Moreover, an additional {\it hardening} is present at $\sim300\ \rm{GeV/n}$  as measured by PAMELA \citep{2011Sci...332...69A}, AMS-02 \citep{2015PhRvL.114q1103A} and CREAM \citep{2010ApJ...714L..89A}. This second feature reproduces the observed MILAGRO anomaly at $15\ \rm{TeV}$ \citep{2015ApJ...815L..25G}. 
In the discussion is also proposed a view of the background model without the additional hardening at $300\ \rm{GeV/n}$  ({\it Gamma Model without hardening}).

The third model proposed is the so-called {\it Base Model}: the diffusion coefficient is considered constant in the whole Galaxy, and corresponding to the value observed at the Earth position ($\Gamma\sim2.6\div2.7$ \cite{2018PhRvD..98c0001T}). Even in this case, the hardening at $300\ \rm{GeV/n}$ is considered. The last representation is the {\it Conventional} description of the CR diffusion.

The obtained CR distributions (with \texttt{DRAGON}) are included as inputs of \texttt{GAMMASKY} code. For modeling the gas structure and distribution of the inner ring of Our Galaxy, an analytical model is used as described in \cite{2007A&A...467..611F}. This model represents a source of systematics since the corresponding distribution of the gaseous content is smooth: there is not a detailed description of the clouds and complex structures filling the CMZ.

Since the observations of the inner ring of our Galaxy are affected by large uncertainties and systematics, the reported hypothesis represents an  extension at the GC position of the trend tuned on local data, and an extrapolation at higher energies. 
A development of {\it Gamma Model} is ongoing, tuned on recent observations and featured by updated maps of the gas distribution in the Galaxy. The next released large-scale models of $\gamma$-ray emission from the whole Milky Way will be released with the use of \texttt{DRAGON2} \citep{2017JCAP...02..015E} and \texttt{HERMES} codes\footnote{\url{https://github.com/cosmicrays/}}. These tools are usable to the scientific community.

\section{Results}
The models described above are compared with the observed data by currently operating gamma-ray observatories. In particular, a detailed analysis of $\sim10$ years of Pass8 \citep{bruel2018fermilat} {\it Fermi}-LAT data has been performed with the latest released IRFs (\texttt{P8R3\_CLEAN\_V2}), isotropic background (\texttt{iso\_P8R3\_CLEAN\_V2}), and 4FGL catalog \citep{Abdollahi_2020}.

\begin{figure}[ht]
  \begin{minipage}[b]{0.5\linewidth}
  \includegraphics[width=1.\linewidth]{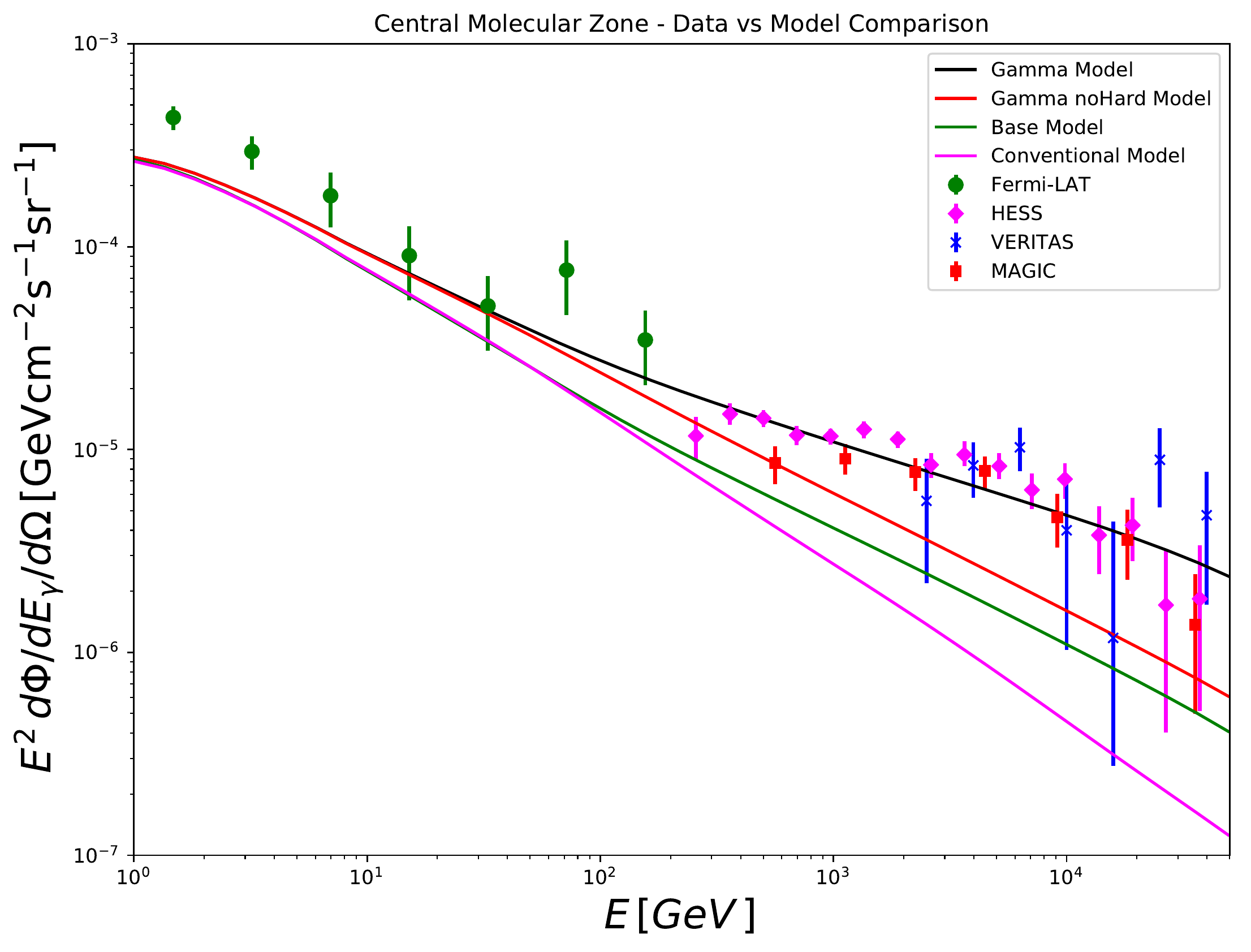}
  \end{minipage} 
  \begin{minipage}[b]{0.5\linewidth}
  \includegraphics[width=1.\linewidth]{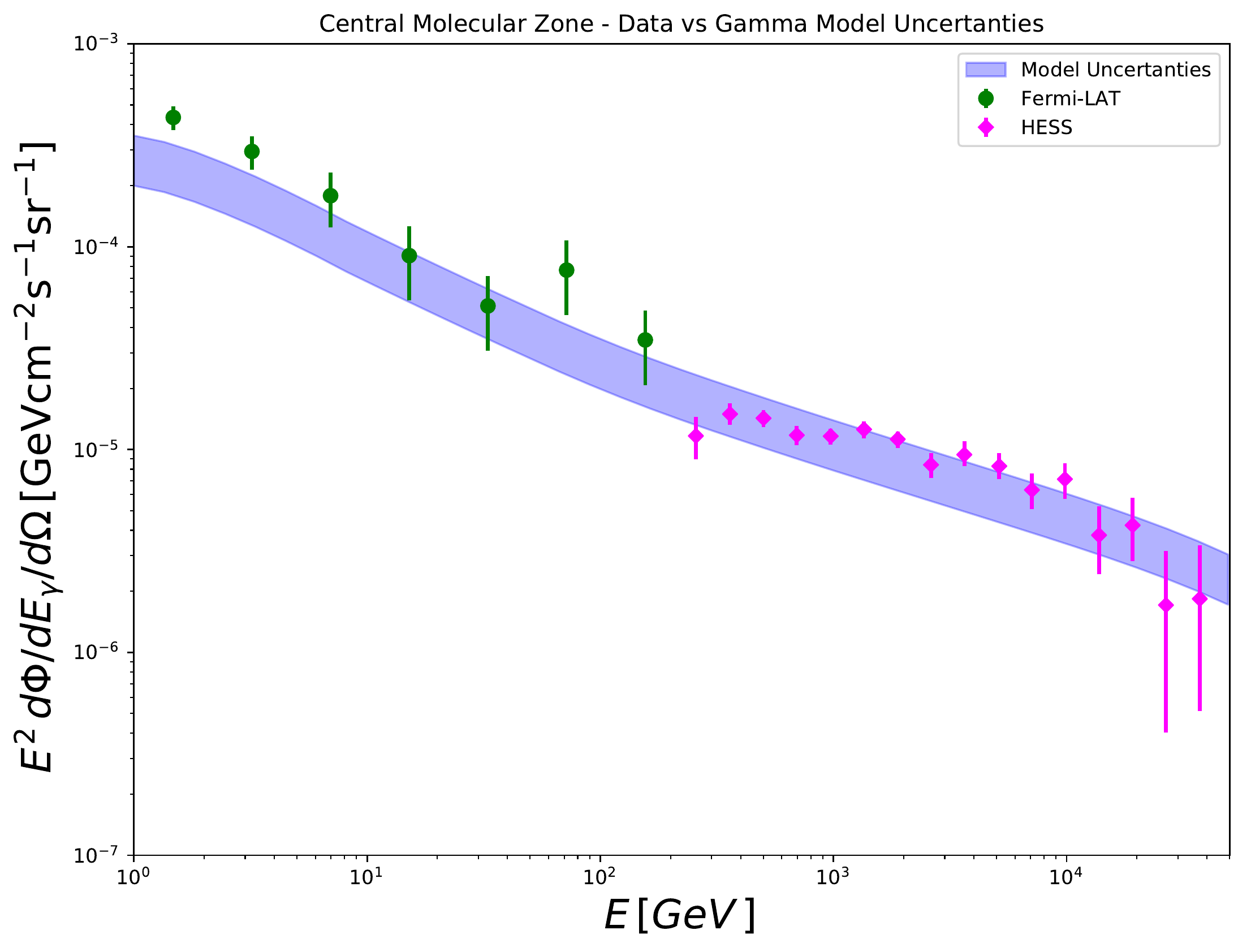}
  \end{minipage}
  \caption{{\it Left}: Compared energy spectra of $\gamma$-ray diffuse emission from the Galactic Ridge with {\it Fermi}-LAT, H.E.S.S. \citep{2018A&A...612A...9H}, MAGIC \citep{2020A&A...642A.190M} and VERITAS \citep{2021ApJ...913..115A} data. {\it Right}: {\it Gamma Model} energy spectrum compared with {\it Fermi}-LAT and H.E.S.S. data in the Galactic Ridge region. The blue band represents the systematic error due to the uncertainty in the estimation of the $X_{\rm{CO}}$ factor at the GC position. }
  \label{compmod}
\end{figure}

In figure \ref{compmod} {\it left} is shown the energy spectrum from $1\ \rm{GeV}$ to more than $50\ \rm{TeV}$ of the $\gamma$-ray emission from the inner part of the CMZ, the {\it Galactic Ridge} (for more details see \citep{2018A&A...612A...9H}).

In this work the analyzed {\it Fermi}-LAT data is compared with H.E.S.S. \citep{2018A&A...612A...9H}, MAGIC \citep{2020A&A...642A.190M} and VERITAS \citep{2021ApJ...913..115A} measurements. Moreover the expected diffuse $\gamma$-ray emission of the four phenomenological models described in the previous section is shown. 

At the {\it Fermi}-LAT energies the models are able to reproduce the observed {\it excess} in the Galactic Ridge, while at higher energies -- essentially at that energies reachable by Cherenkov Telescopes -- the four energy spectra reproduce a different behaviour of the modelled $\gamma$-ray diffuse emission. 
In particular the measured emission is reproduced by the {\it Gamma Model}, but not properly by the other three models.

Instead, in figure \ref{compmod} {\it right}, the measured {\it Fermi}-LAT and H.E.S.S. data are  compared with the expected $\gamma$-ray emission computed with {\it Gamma Model}. 
The blue band represents the uncertainty in the normalization of the spectrum of the diffuse emission. Indeed in \cite{2007A&A...467..611F} for estimating the gas mass of the inner ring of Our Galaxy, has been considered the so-called $X_{\rm{CO}}$ factor \citep{1975ApJ...202...50D} equals to $\sim(2\div4)\times10^{19}\ \rm{cm}^{-2}\rm{K}^{-1}\rm{km}^{-1}\rm{s}$ with an uncertainty of order $2$.  This causes a corresponding uncertainty in the prediction of the model used to compute the expected $\gamma$-ray emission from the region.
 %
%

\section{Discussion and Conclusions}
In the H.E.S.S. Collaboration papers \citep{2016Natur.531..476H, 2018A&A...612A...9H} the observed $\gamma$-ray {\it excess} has been interpreted in terms of the so-called {\it PeVatron} scenario. The inferred CR density is compatible with CR escaping from a {\it steady-state} source that continuously injects CRs in the region. The presence of a PeVatron is associated with the Super Massive Black Hole (SMBH) Sgr A$^\star$ in the center of the Milky Way, or with a hidden accelerator located in its vicinity, for instance associated with a compact {\it Stellar Wind Cluster} \cite{2019NatAs...3..561A, 2021MNRAS.504.6096M}.

In this paper instead, the observed flux from the Galactic Ridge has been interpreted in terms of the {\it hard} component of the diffuse $\gamma$-ray emission featured by the radial dependence of the diffusion coefficient, and an {\it hardening} at $300\ \rm{GeV/n}$. 
As mentioned above, this hypothesis represents an extension at the GC position of the behaviour observed between $8\ \rm{kpc}$ and $3\ \rm{kpc}$, and tuned on local data. Moreover, the extrapolation at higher energies -- the energies reachable by IACTs -- is strongly dependent on the CR transport parameters variation (figure \ref{compmod} {\it left}).

Both {\it PeVatron} scenario and {\it hard-diffusion} scenario are able to reproduce the observed spectrum in the innermost part of Our Own Galaxy: with the currently operating $\gamma$-ray observatories it is not possible to know with sufficient accuracy the origin and the nature of the observed $\gamma$-ray excess, and to reach definitive conclusions. 

The opportunity to clarify the intricate panorama featuring the center of the Milky Way is given by the observation of different regions farther from the GC. For instance, the recently discovered HESS J1741-302 \citep{2018A&A...612A..13H} may represent an ideal target for testing the impact of different scenarios (for more details see \cite{2019ICRC...36..816V}). 
Another interesting region is represented by {\it Bania Clump}, a complex, dense and giant molecular cloud located at $\sim500\div800\ \rm{pc}$ away from the GC (see \cite{2018VENTURA}). 
Indeed these kind of objects represent ideal targets for CR hadrons interaction and subsequently $\gamma$-rays production. Whether the presence of potential PeVatron sources is sufficiently far way from selected peculiar dense clouds, the well-know $1/r$ CR radial profile is not able to {\it shine} these regions, but the level of the $\gamma$-ray {\it background} emission featured by the {\it hardening}, described above, may be high enough to illuminate dense clouds. In this context, an accurate 3D description of the morphology and dynamics of the clouds filling the inner part of Our Galaxy is fundamental and required.

The opportunity to disentangle among the several suggested scenarios will be given by the upcoming Cherenkov Telescope Array (CTA) thanks to the increased sensitivity and angular resolution. At that point, the nowadays {\it dark regions} -- too faint to be detected by currently operating $\gamma$-ray observatories -- may be bright enough to be detected by CTA, and shed light on the origin of this peculiar emission. 
Moreover the detection of a {\it cut-off} in the observed energy spectrum will give the definitive interpretation on the nature and origin of the emission independently by models \citep{2020ApJ...903...61C}.

\bibliographystyle{JHEP}
\bibliography{skeleton.bib}


%
%
%

\end{document}